\documentclass[12pt]{article}
\usepackage{amsmath}
\usepackage{amssymb}
\usepackage{graphicx}
\usepackage{natbib}
\hyphenchar\font=-1
\usepackage[none]{hyphenat}
\usepackage{authblk}

\begin{document}

\title{A Quantitative Test of Population Genetics Using Spatio-Genetic Patterns in Bacterial Colonies}
\author{K.~S.~Korolev} 
\affil{Department of Physics, Massachusetts Institute of Technology, Cambridge, Massachusetts 02139, USA; papers.korolev@gmail.com}
\author{Jo\~ao~B.~Xavier} 
\affil{Program in Computational Biology, Memorial Sloan-Kettering Cancer Center, New York, New York 10065, USA; XavierJ@mskcc.org}
\author{David~R.~Nelson}
\affil{Department of Physics and FAS Center for Systems Biology, Harvard University, Cambridge, Massachusetts 02138, USA; nelson@physics.harvard.edu}
\author{Kevin~R.~Foster}
\affil{FAS Center for Systems Biology, Harvard University, Cambridge, Massachusetts 02138, USA; Department of Zoology, Oxford University, South Parks Road, Oxford, OX1 3PS, UK; Oxford Centre for Integrative Systems Biology, Oxford University, South Parks Road, Oxford, OX1 3QU, UK; kevin.foster@zoo.ox.ac.uk}

\date{}
\maketitle
{This manuscript is an article.}

\newpage

\begin{abstract}
It is widely accepted that population genetics theory is the cornerstone of evolutionary analyses. Empirical tests of the theory, however, are challenging because of the complex relationships between space, dispersal, and evolution. Critically, we lack quantitative validation of the spatial models of population genetics. Here we combine analytics, on and off-lattice simulations, and experiments with bacteria to perform quantitative tests of the theory.  We study two bacterial species, the gut microbe \textit{Escherichia coli} and the opportunistic pathogen \textit{Pseudomonas aeruginosa}, and show that spatio-genetic patterns in colony biofilms of both species are accurately described by an extension of the one-dimensional stepping-stone model. We use one empirical measure, genetic diversity at the colony periphery, to parameterize our models and show that we can then accurately predict another key variable: the degree of short-range cell migration along an edge. Moreover, the model allows us to estimate other key parameters including effective population size~(density) at the expansion frontier. While our experimental system is a simplification of natural microbial community, we argue it is a proof of principle that the spatial models of population genetics can quantitatively capture organismal evolution. 
\end{abstract}

\newpage

\section*{Introduction}
\label{SIntroduction}

The neo-darwinian synthesis brought population genetics into the heart of evolutionary biology~\citep{Wright:Landscape,Fisher:Selection,Dobzhansky:Genetics}, something that has since been crystallized in the vast array of sophisticated models, which incorporate factors like stochasticity, spatial structure, and epistasis~\citep{Crow:PopulationGenetics,Felsenstein:Review,Loewe:Review}. Despite this and the widespread acceptance of the population genetics as the cornerstone of the evolutionary theory, the low ratio of empiricism to modeling has long been noted in the literature~\citep{Mayr:Where,Rousset:Book}. Although genomics studies can be a very useful for testing qualitative predictions of population genetics, e.g. the importance of effective population size for purifying selection~\citep{Lynch:Intron}, a controlled experiment in the laboratory is better suited for the specific goal of validating the theory.

One of the key barriers to quantitative tests is that many theories of population genetics consider well-mixed populations, i.e. populations without spatial structure. The assumption of a lack of structure can often be a drastic approximation. Moreover, the ever-growing number of theories that do incorporate spatial structure~\citep{Excoffier:Review,Rousset:Book,nagylaki:tutorial} presents a new set of problems because empirical tests require the knowledge of certain demographic parameters, such as dispersal rates, for which little data are typically available~\citep{Rousset:Book}. An open challenge for population genetics, therefore, is to develop new empirical methods and tests of the theories themselves, and, in particular, quantitative tests that can rigorously verify models of spatially structured populations. 

We focus on neutral population genetics theory of effectively one-dimensional populations applied to range expansions. The importance of spatial separation for the development of genetic diversity was first emphasized by~\cite{Wright:IsolationByDistance} and was further developed by~\cite{Malecot:DecreaseReltationshipDistance,Malecot:Dynamics}; see also \citep{KimuraWeiss:SSM,Nagylaki:DecayGeneticVariability,Barton:NeutralEvolution}. In addition to stationary populations, Mal{\'e}cot's  analysis also describes populations undergoing range expansions as was shown in \citep{Korolev:Review}. While the theory is in principle general to a range expansion of any species, we focus here on one of the most tractable of experimental systems, the colony biofilms formed by bacteria on solid surfaces~\citep{rani:spatial_heterogeneity}.

Microbial populations are often used to elucidate the mechanisms of evolution because of their short generation times and the ease of genetic manipulation~\citep{Elena:ExperimentalEvolutionReview,Barrett:EvolutionComplexEnvironment,West:Social}. With a few exceptions~\citep{rainey:radiation,kerr:tragedy,rainey:conflict}, the homogeneous environment of a chemostat or a vigorously shaken test tube is preferred because of its simplicity and the ease of comparison to mathematical models, which are often intractable for populations with spatial structure. However, even for microbes, it is now clear that most natural populations are not well-mixed, and the neighborhood of any individual is enriched in the individuals that are ancestrally related to it. This is especially true for many bacteria commonly forming dense surface-attached communities known as biofilms~\citep{kolter:superficial_life,Nadell:Sociobiology}. These resilient microbial communities are found in numerous contexts, both medical and industrial, and are often highly resistant to antibiotics. For this reason, there is now a large body of literature on biofilm formation, which underlines their complexity with major differences in the regulation and development of biofilms among strains and species~\citep{beloin:biofilm_expression}. Key factors affecting the structure of biofilms include physical interactions among cells, the diffusivity of metabolites~\citep{dockery:biofilm_fingers,stewart:diffusion}, and the genetic regulation of growth rate in response to both nutrient levels~\citep{vulic:cheating} and quorum sensing~\citep{parsek:quorum,vulic:cheating,hammer:quorum,nadell:quorum}. Growth in microbial groups also often relies on growth-promoting secretions, which themselves can be under nutrient and quorum sensing regulation~\citep{xavier:secretions}.

Despite the potential complexities, the spatio-genetic patterns in microbial biofilms can be visualized by fluorescent labeling part of the population~\citep{klausen:biofilm,HallatschekNelson:ExperimentalSegregation}. With this setup, cells of the same color can be ancestrally related, while cells of different colors cannot (provided mutations that switch colors can be neglected). Colony biofilms are particularly amenable to such visualizations~\citep{rani:spatial_heterogeneity}.  As an example, the fluorescent image of a bacterial colony is shown in Fig.~\ref{fig:illustrations}A. The most striking feature of this image is that an initially well-mixed population of two alleles spatially separates or demixes into monoallelic sectors. The sectoring pattern in this colony is due to genetic drift at the front of the expansion, i.e. stochastic changes in the allele frequencies due to random variation in the number of offspring keeping up with the outgoing expansion. Since at any given time only a few cells at the edge of the expansion colonize the Petri dish, the fluctuations can be large and lead to random local increase in the number of cells of one color at the expense of the other. Once one of the colors becomes extinct as a result of this stochastic process in a small region of the expanding colony, a sector is formed, and it persists as long as its boundaries do not coalesce. Such spatio-genetic correlations significantly affect the evolutionary dynamics of populations~\citep{Korolev:Review}. General examples range from the neutral theory in ecology~\citep{Houchmandzadeh:NeutralEcology} and evolution~\citep{KimuraWeiss:SSM,Charlesworth:NeutralVariation}, through the propensity for malignant tumor formation~\citep{Michor:Cancer}, to the evolution of cooperation~\citep{Nowak:SpatialPD92,Nowak:SpatialPD94,Hauert:SpatialSnowDrift}. Indeed, the latter has direct relevance to microbes and the evolution of growth-promoting secretions that are cooperative in the sense that the secretion of one cell will help others grow. These secretions often require strong spatio-genetic structure that prevent exploitative genotypes from using the secretions of the cooperative genotypes~\citep{Nadell:EmergenceSpatialStructure,Nadell:Sociobiology}.

Here, we use the spatio-genetic patterns in bacterial colonies to test the population genetics models of range expansions. We begin by synthesizing and extending the theory of range expansions in a way that allows us to compare the predictions of this theory to our experimental data. In particular, two aspects of the theory are novel. First, we find that chiral or non-neutral growth results in sector boundaries being logarithmic spirals [Eq.~(\ref{EAveragePhi})]. Second, we present a detailed discussion on the role of population density on the number of surviving sectors. This theory of range expansions is tested using experiments with two ecologically distinct bacterial species: the gut microbe \textit{Escherichia coli} and the opportunistic pathogen \textit{Pseudomonas aeruginosa}. Specifically, we test the dependence of the genetic diversity at the population front~(sector number) on the physical size of the founding population~(initial radius of the colony) and check the assumption that boundaries between genotypes perform radial random walks. We show that the model not only captures the qualitative relationship between demographic parameters and spatio-genetic patterns, but it also allows us to estimate these parameters with considerable accuracy. This quantitative validation of the theory supports the utility of population genetics models to capture the evolutionary process.

\section*{Methods}
\label{SM}

Data deposited at Dryad: doi:10.5061/dryad.3147q.

\subsection*{Bacterial Strains and Plasmids}

We used the same two strains of \textit{E. coli} as in~\citep{HallatschekNelson:ExperimentalSegregation}, where it was shown that these strains are equally fit in well-mixed liquid culture and genetically identical except for the different fluorescent protein gene carried on plasmids. The \textit{P. aeruginosa} strain was generously provided by Roberto Kolter and was originally isolated from fresh water in Columbia. We labeled this strain with cfp and yfp~\citep{Xavier:SpatialBiofilm} resulting in two strains, which are also genetically identical except for the fluorescence gene inserted in the chromosome. The pairs of colored strains within each species grew at the same rate and behaved neutrally on the plates. This was confirmed from the shape of the sector boundaries in the experiments~\citep{Hallatschek:LifeFront}. In particular, strains with a fitness advantage would tend to curve outward and cut off neighboring strains; see Eq. (9). This boundary bending was not observed and the analysis of sector boundaries showed that, both for \textit{E. coli} and \textit{P. aeruginosa}, the pairs of colored strains had the same fitness in our experiments; see Results. 

\subsection*{Growth Conditions}

Each strain was first grown separately in~$3\;\rm{mL}$ of liquid LB medium supplemented with the appropriate antibiotic~(carbenicillin for \textit{E. coli} and gentamycin for \textit{P. aeruginosa}) at the concentration of~$100\;\mu \rm{g}/\rm{mL}$. The test tubes were incubated at~$37^{\circ}\rm{C}$ and constantly shaken. After an overnight growth, bacteria were diluted to an optical density of $0.05$ (at $600\;\rm{nm}$ wavelength) in~$3\;\rm{mL}$ of liquid LB medium without antibiotics and grown for $2$~hours. Then, the two strains of interest were mixed in 1:1 ratio, as measured by optical density, and a small volume of bacterial mix ranging from $1\;\mu\rm{L}$ to $64\;\mu\rm{L}$ was pipetted on a LB-agar plate with~$1.5\%$ agar and~$100\;\mu \rm{g}/\rm{mL}$ of appropriate antibiotic. The plates were then incubated at~$21^{\circ}\rm{C}$ or~$37^{\circ}\rm{C}$ and high humidity. \textit{E. coli} strains did not grow at~$21^{\circ}\rm{C}$.

\subsection*{Microscopy and Image Processing}

Fluorescent images were obtained with a fluorescence stereoscope Zeiss Lumar V.12 and a scanner Amersham 392 Typhoon 9400 (GE Healthcare). Scanned plates were imaged from the bottom at~$50\;\mu\rm{m}$ resolution with Alexa 526 filter. 

The initial radii of the colonies were measured within an hour of inoculation by fitting a circle to the colony boundary using the fluorescent microscope software. The number of sectors in each colony was calculated manually. \texttt{MatLab R2007} was used to extract the radii of the colonies at later times and the shape of sector boundaries from the images. In particular, we used the routine~\texttt{edge} to identify sector boundaries, and a least square fitting routine to find the circumference of colonies~\citep{Pratt:CircleFit}.

To calculate the heterozygosity~$H(t,\phi)$ defined by Eq.~(\ref{eq:heterozygosity}) and shown in Fig.~\ref{fig:heterozygosity}, we obtained the allele frequencies~$f(t,\phi)$ along a set of concentric circles~(which stand for different time points). While, in simulations, the frequencies are known, we had to calculate~$f(t,\phi)$ in experiments from the fluorescence intensities of the colony images. This was done by choosing a threshold intensity such that any pixel with higher intensity is assigned~$f=1$ and any pixel with lower intensity is assigned~$f=0$. We then used Eq.~(\ref{eq:heterozygosity}) to calculate the heterozygosity, and the averaging was done over all the points angle~$\phi$ apart in a given circle and all the circles corresponding to the same time point in the colonies used in that experiment.     

\subsection*{Estimating~$D_{g}$ and~$D_{s}$}

Here we describe how~$D_{g}$ and~$D_{s}$ were estimated from fluorescent images of genetic demixing. The diffusion constant~($D_{s}^{m}$ in Table~\ref{TNRFit}) was estimated from the random walks of sector boundaries. We first obtained the polar coordinates~$\varphi(r)$ of individual sector boundaries as described above. Once the coordinates were known for~$r\in(r_{i},r_{f})$, we evaluated~$\varphi(r)-\varphi(r_{i})$ for each boundary. We then estimated the mean and variance of this quantity as a function of~$\ln(r/r_{i})$ and~$1/r_{i}-1/r$ respectively. The model accurately described boundary motion, and the variance of~$\varphi(r)-\varphi(r_{i})$ was a linear function of~$1/r_{i}-1/r$; therefore,~$2D_{s}/v_{\parallel}$ was found from the least mean square estimate of the slope.

In addition, both~$D_{g}$ and~$D_{s}$~($D_{s}^{e}$ in Table~\ref{TNRFit}) were estimated from the experiments done with different volumes of inoculant. Different volumes of inoculant led to different initial radii of the colonies, which were measured as described above. The observed values of radii were then binned, and the average number of sectors at the end of the experiment (when coalescence of sectors ceased) were calculated within each bin. The average number of sectors was a linear function of the square root of the initial radius. The least mean square estimate equals~$H_{0}\sqrt{2\pi v_{\parallel}/D_{s}}$ for the slope and~$2\pi H_{0}v_{\parallel}/D_{g}$ for the intercept; see Eq.~(\ref{ENR}). Since both~$H_{0}$ and~$v_{\parallel}$ were measured directly, these least mean square estimates allowed us to measure~$D_{g}$ and~$D_{s}$. 

\subsection*{On-Lattice Simulations}

We also tested our genetic demixing model described below via simulations designed to mimic range expansions of natural populations. In these simulations, the habitat was a square lattice of islands that could support at most~$N$ individuals, similar to that in the stepping-stone model~\citep{KimuraWeiss:SSM}. During every time step, each island had one opportunity for outward migration and one opportunity for growth.

During the migration update, an individual from an island could migrate to a randomly selected nearest neighbor island with a probability proportional to the fraction of cells on the island and the fraction of empty spots~(vacancies) on that neighbor island. The coefficient of proportionality~$d\le1$ determined the magnitude of the effective diffusion constant, and was used to vary the rate of migration. Since the migration rate was proportional to the fraction of vacancies, an organism could exchange positions only with a vacancy, not another organism. Hence, there was no migration in the dense-packed interior of the simulated colony. Its spatio-genetic pattern was thus a frozen record of the genetic composition of the expansion front, similar to  ~\citep{HallatschekNelson:ExperimentalSegregation}. Hence, we analyzed both the experimental data and the simulation data on an equal footing. Periodic boundary conditions were used on all four sides of the two-dimensional habitat.

During the growth update, a randomly selected individual in each island reproduced with a probability proportional to the fraction of the cells on the island and the fraction of vacancies on that island; the coefficient of proportionality~$p\le1$ allowed us to vary the growth rate. We checked that this update rule could reproduce simple logistic growth~\citep{Murray:MathematicalBiology}. To compare with the experiments, we simulated only two alleles and tracked the genetic identity of the individuals throughout the simulation. A typical simulated range expansion is shown in Fig.~\ref{fig:illustrations}B.

\section*{Off-Lattice Simulations}

We also carried out simulations using an off-lattice, agent-based approach that explicitly described the dynamics of each agent based on the micro-environment it perceived. The agents represented cells and their behavior was governed by the rules that mimic the behavior of real bacteria: they increased in radius when taking up nutrients, divided once they reached a certain critical size, and moved continuously~(off-lattice) when pushed by neighboring cells. The pushing algorithm was based on a hard-sphere model where each agent moved in the direction minimizing the overlap with neighboring agents. The model used a fast numerical method (FAS multigrid) to solve partial differential equations for the nutrient concentration. The algorithms were described in a previous publication~\citep{Xavier:Algorithm}.

We simulated neutral evolution experiments by implementing two genotypes with equal phenotypes except for their color. Agent growth was determined by the local concentration of the growth-limiting nutrient following Monod kinetics, see  ~\citep{Nadell:EmergenceSpatialStructure}. Simulations were initiated with a number of individuals placed in a circular homeland. The growth of the agents and colony was then followed in time, producing structures such as the one shown in Fig.~\ref{fig:illustrations}C.

\section*{Models of range expansions}
\label{STP}

The importance of space to evolutionary processes has been studied in many disciplines including
population genetics, ecology, and probability theory, where the classic spatial models, for example the stepping-stone model~\citep{KimuraWeiss:SSM,Nagylaki:DecayGeneticVariability,Malecot:Dynamics}, Hubbell's model~\citep{Hubbell:Book,Houchmandzadeh:NeutralEcology}, and the voter model~\citep{CoxGriffeath:VoterModel} have been developed for stationary~(non-expanding) populations.
Following \citep{Korolev:Review}, we use coalescent theory~\citep{Kingman:Coalescent} to track the genetic diversity at different points in space and adapt the classic results from spatial population genetics~\citep{Malecot:DecreaseReltationshipDistance,Malecot:Dynamics,KimuraWeiss:SSM,Nagylaki:DecayGeneticVariability,Barton:NeutralEvolution} to understand the evolutionary dynamics of clonal populations during a range expansion.

While neutral expansions typically occur on a two-dimensional surface, a one-dimensional model is sufficient to understand how genetic diversity changes at the front of the expanding population. For stationary (non-expanding) populations, the dimensionality of space plays an important role~\citep{Barton:NeutralEvolution,Korolev:Review}, and dimensional reduction from two to one spatial dimensions is impossible. In expanding populations however, this dimensional reduction is possible because all relevant dynamics occurs at the perimeter of the population, which is effectively one-dimensional. Indeed, the organisms far behind the expansion front cannot affect the future genetic composition at the frontier because their offspring are not part of the outgoing expansion wave. Although the dynamics of allele frequencies at the front can be described by an effective one-dimensional model, this is not in general true for the population in the wake of the expansion, for which a full two-dimensional model is necessary. Nevertheless, the results of the one-dimensional model are applicable behind the front on the time scales when migration in the interior can be neglected~\citep{Hallatschek:LifeFront}.  

Although a one-dimensional model is easier to analyze, the demographic parameters in this model are in general complicated functions of the demographic parameters of the original two-dimensional population. For example, the effective linear density~(carefully defined in Appendix~A) of the one-dimensional population at the frontier depends on the carrying capacity of the habitat, the shape of the density profile at the front, and the probability of individuals slightly behind the expansion edge to migrate forward\footnote{Organisms expanding in spatially homogeneous habitats typically reach the carrying capacity in the interior of the population, but the population density decays to zero beyond the expansion front. Here, we assume that the density profile at the front remains constant during the expansion resulting in a constant population density in the effective one-dimensional model. This behavior is typical for reaction-diffusion systems with saturation~\protect{\citep{Fisher:FisherWave,Kolmogorov:FKPPEquation,Murray:MathematicalBiology}} and is confirmed in our simulations; see also \protect{\citep{Nadell:EmergenceSpatialStructure}}.}~\citep{Hallatschek:1dWave}. The exact nature of these dependences varies for different reproduction and migration models, but the effective one-dimensional description is universal, once model-specific details are absorbed in the effective demographic parameters.

To make a direct connection with the subsequent experiments, we assume that only two genotypes are present, and that they have the same fitness. We also assume that the colony radius~$R$ increases linearly with time~$t$:

\begin{equation}
R(t)=R_{0}+v_{\parallel}t,
\label{ERt}
\end{equation}

\noindent where~$v_{\parallel}$ is the radial expansion velocity, and~$R_{0}$ is the radius of the colony when~$t=0$. Expansions with a constant velocity are expected for reaction-diffusion population dynamics~\citep{Murray:MathematicalBiology} and have been observed in microbial colonies~\citep{HallatschekNelson:ExperimentalSegregation,Wakita:Expansion}. Range expansions in our simulations also obey Eq.~(\ref{ERt}).

\begin{figure}[!ht]
\begin{center}
\includegraphics[width=12.46cm]{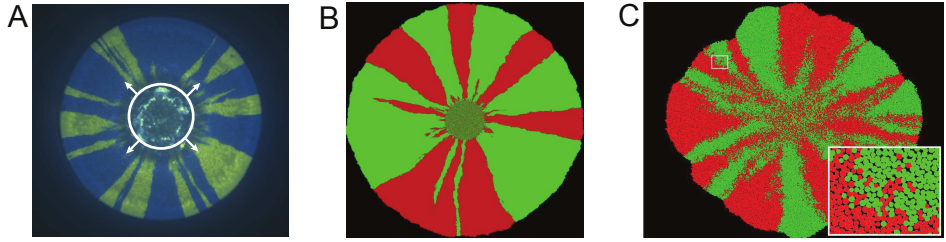}
\caption{\small Genetic demixing in experiments~(A), on-lattice simulations~(B), and off-lattice simulations~(C). Different colors label different alleles, and the initial mixing ratio was 1:1. (A) The Petri dish was inoculated with a drop~(bounded by the white circle) of a mixture of \textit{Pseudomonas aeruginosa} cells labeled with two different fluorescent markers that do not affect the relative fitness of the cells. As the drop dries out and the colony starts to expand~(shown with arrows), the population at the front of the expansion demixes~(separates) into sectors of different colors. This genetic demixing is due to the loss of local genetic diversity during the expansion, see \protect{\citep{HallatschekNelson:ExperimentalSegregation,Hallatschek:LifeFront,Korolev:Review}}. This plate was incubated at~$37^\circ{\rm C}$. (B) The habitat was a~$1000\times1000$ array of islands arranged on a square lattice. Each island could harbor at most~$N=30$ individuals, consistent with the width of the layer of actively growing cells found in~\protect{\citep{HallatschekNelson:ExperimentalSegregation}}. The simulation was started with a well-mixed population occupying a disk of radius~$R_{0}=80$ lattice spacings in the center of the habitat. (C) The simulation was initiated with~$1024$ agents placed at random within a circular inoculant in the center of the habitat. The boxed region of the colony is shown in the inset at a higher magnification. We attribute the slightly square shape of the colony to the underlying square grid used to solve the nutrient reaction-diffusion equation.}
\label{fig:illustrations}
\end{center}
\end{figure}

Spatial genetic demixing~(Fig.~\ref{fig:illustrations}) is one of the key predictions of this model. Although the total number of individuals is very large, locally only a finite number reproduce, leading to genetic drift (i.e. fluctuations in the genetic composition of a population due to the random sampling of offspring to form next generation). Over time, these demographic fluctuations tend to reduce genetic diversity at the growing front of the population. Eventually, a single genotype will reach fixation at any point on the expanding edge. In bacterial colonies, this fixation manifests as the formation of a sector~(Fig.~\ref{fig:illustrations}A).

Below we present two complimentary ways to rigorously analyze the behavior of the one-dimensional population at the expansion front. The first approach is an extension of classic spatial population genetics~\citep{Malecot:DecreaseReltationshipDistance,Malecot:Dynamics,KimuraWeiss:SSM,Nagylaki:DecayGeneticVariability,nagylaki:sites,Barton:NeutralEvolution} using coalescent theory~\citep{Kingman:Coalescent} to track the genetic diversity at different points in space. At the core of this approach is the stepping-stone model~\cite{KimuraWeiss:SSM}, which describes migration and reproduction of individuals in a population. A concise description of this model is given in Appendix~A. The second modeling approach was proposed in~\citep{HallatschekNelson:ExperimentalSegregation,Hallatschek:LifeFront} and tracks the domain \textit{boundaries} between the regions with different genotypes. While it is applicable only after monoallelic domains~(sectors) form, the second approach yields predictions complimentary to the predictions of the first approach. We use these additional predictions to make an quantitative experimental test of our theory.

\subsection*{Model 1: Tracking local genetic diversity}

Spatial genetic demixing can be most readily understood by calculating the genetic diversity of the population over time. The genetic composition of the front at time~$t$ is fully described by the relative fraction~(frequency) of one of the two alleles~$f(t,\phi)$, where~$\phi$ is the azimuthal angle around the center of the population. The relative fraction of the other allele is then~$1-f(t,\phi)$. These fractions change in time because of migration and reproduction. One can think of the population as a circle of islands~(demes) each carrying a certain number of organisms~(this number of organisms sets the strength of genetic drift). Each generation, the organisms reproduce within a deme, say by a Wright-Fisher process, and migrate between neighboring demes. This conceptual population model is the stepping-stone model of~\cite{KimuraWeiss:SSM}; we present a mathematical formulation of the stepping-stone model in Appendix~A.

Because~$f(t,\phi)$ is a random variable, which is different for every expansion, we need to calculate quantities averaged over repeated expansions to compare the predictions of the model to experimental data. The first moment~(mean) of~$f(t,\phi)$ does not contain information about genetic demixing, because the relative fraction of two neutral genotypes does not change on average

\begin{equation}
 \mathbb{E}[f(t,\phi)]  = f_{0},
\end{equation}
 
\noindent where~$f_{0}$ is the initial frequency of allele one, and~$\mathbb{E}$ denotes averaging with respect to different realizations of the range expansion.

The second moment~(variance), however, captures the spatio-genetic correlations. We define the average spatial heterozygosity as

\begin{equation}
\begin{split}
H(t,\phi_{1}-\phi_{2}) =&  \mathbb{E}\{f(t,\phi_{1})[1-f(t,\phi_{2})]\} \\& + \mathbb{E}\{f(t,\phi_{2})[1-f(t,\phi_{1})]\}, 
\end{split}
\label{eq:heterozygosity}
\end{equation}

\noindent which is the probability of sampling two different alleles separated by distance~$|\phi_{1}-\phi_{2}| R(t)$ along the circular frontier. Note that~$H$ depends only on the relative separation between the two points because of rotational invariance of the averages.

\begin{figure}[!ht]
\begin{center}
\includegraphics[width=14 cm]{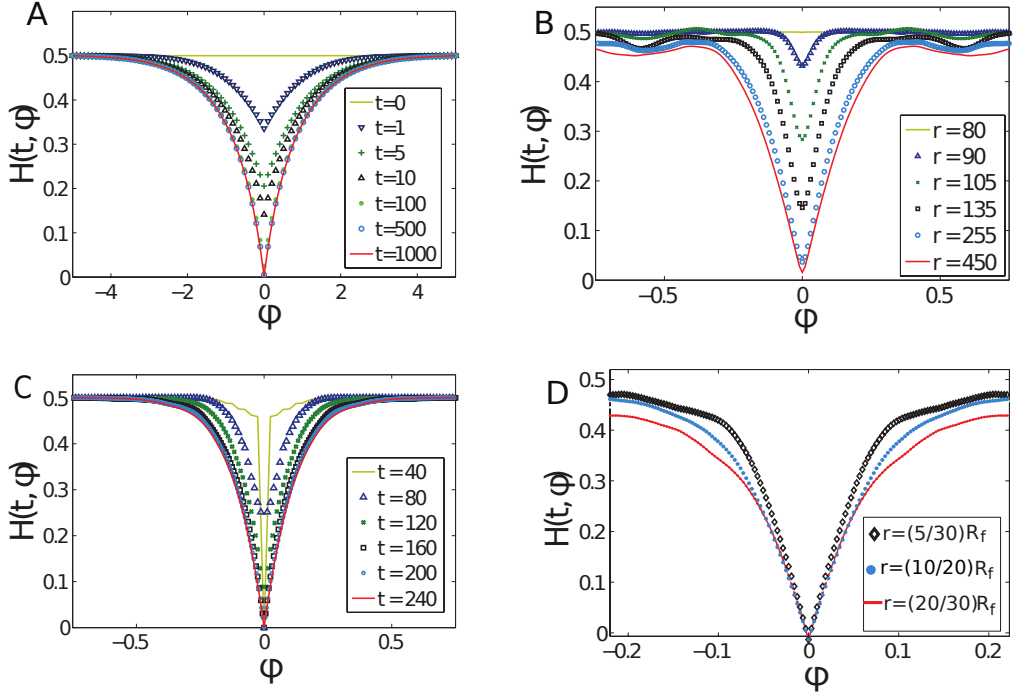}
\caption{\small Plots of average spatial heterozygosity as the function of angle during range expansions from 1:1 well-mixed populations in our model (A), on-lattice simulations (B), off-lattice simulations (C), and experiments (D). (A) Solution of Eq.~(\ref{EOMH}) with~$D_{s}=1$, $D_{g}=1$, $v_{\parallel}=1$, and $R_{0}=1$ at various times~$t$. Note that there is no significant difference between~$H(500,\phi)$ and~$H(1000,\phi)$ because~$H(t,\phi)$ reaches a nontrivial limit-shape as~$t\rightarrow\infty$. (B) $H(t,\phi)$ from 24 on-lattice simulations with the same parameters as in Fig.~\protect{\ref{fig:illustrations}}B, except~$N=300$. In agreement with (A), we see a gradual decrease of~$H(t,0)$ with time. The radius~$r=v_{\parallel}t+R_{0}$ is in a direct correspondence with time~$t$. (C) $H(t,\phi)$ at the expansion frontier from~$10$ off-lattice simulations, as in Fig.~\protect{\ref{fig:illustrations}}C. Because off-lattice simulations model a monolayer of cells, any spatial point in a colony has a unique genetic state; hence~$H(t,0)=0$. (D) The average spatial heterozygosity was calculated from~$8$ \textit{E. coli} colonies inoculated with~$3\;\mu\rm{l}$ of the bacterial mix of cells. As in~(B), radius~$r$ is directly related to time~$t$ because the spatio-genetic pattern is a frozen record of the genetic composition of the front. The dip in~$H(t,\phi)$ widens with time in agreement with Eq.~(\ref{EOMH}). Similar to (C),~$H(t,0)=0$ because we assume that~$f(t,\phi)$ is either zero or one at every pixel. This assumption is only valid after the initial fixation time discussed in the main text.}
\label{fig:heterozygosity}
\end{center}
\end{figure}

At the beginning of the expansion,~$H(0,\phi)=H_{0}=2f_{0}(1-f_{0})$ because the individuals are well mixed. As the expansion progresses and large monoallelic domains appear, the average spatial heterozygosity decreases for~$\phi$ smaller than the typical domain size because sampling two different alleles in close proximity is unlikely when the population is demixed. For~$\phi$ much larger than the typical domain size, the average spatial heterozygosity is close to~$H_{0}$ because the two sampling points belong to different and uncorrelated sectors carrying allele one with probability~$f_{0}$ or allele two with probability~$1-f_{0}$. Hence, the spatial scale over which~$H(t,\phi)$ increases from its lowest value to~$H_{0}$ characterizes the typical domain size. (See Fig.~\ref{fig:heterozygosity}A and the discussion below.)

As in well-mixed populations, the depression of~$H(t,\phi)$ for small~$\phi$ can be understood by working backward in time and calculating the probability of two individuals to have a common ancestor. Apart from~$t$, this probability depends on the spatial separation of the individuals, the extent of lateral movement during the expansion, and the density of the population. Indeed, for their ancestral lineages to coalesce~(converge on the common ancestor), the spatial random walks performed by the lineages must move to the same spatial location, and the lineages must originate from the same individual in that location. Therefore, the forward-in-time equation for the dynamics of~$H(t,\phi)$ must have a diffusion term to account for the lateral random walk of the ancestral lineages and a sink term to account for the loss of genetic diversity due to lineage coalescence. This equation is rigorously derived in ~\citep{Korolev:Review} (see also Appendix~A) and is given below

\begin{equation}
\label{EOMH}
\begin{split}
\frac{\partial}{\partial t}H(t,\phi)=&\frac{2D_{s}}{(R_{0}+v_{\parallel}t)^{2}}\frac{\partial^{2}}{\partial \phi^{2}}H(t,\phi) \\& -\frac{D_{g}}{R_{0}+v_{\parallel}t}H(t,0)\delta(\phi),
\end{split}
\end{equation}
 
\noindent where~$\delta(\phi)$ is Dirac's delta function,~$D_{s}$ is the spatial diffusion constant of an ancestral lineage, and~$D_{g}\delta(x)$ is the coalescence rate of two ancestral lineages separated by distance~$x$ along the front. This rate and, therefore,~$D_{g}$ are inversely proportional to the generation time and decreases with the population density at the front because the probability of two nearby organisms to be siblings is inversely proportional to the effective population density~\citep{ewens:ne}, defined in Appendix~A. In other words,~$D_{g}$ is a phenomenological parameter characterising the strength of genetic drift. Although one may naively expect that~$D_{g}$ is inversely proportional to the actual population density (or carrying capacity), this is often not the case~\citep{Hallatschek:1dWave}. The factors of~$R_{0}+v_{\parallel}t$ are necessary to relate the angular coordinate~$\phi$ to distances along the constantly inflating circumference of the population. 

From the forward-in-time perspective, the first term on the right hand side of Eq.~(\ref{EOMH}) represents the diffusive mixing in the population, and the second term is responsible for the loss of genetic diversity due to sampling. Smaller population densities lead to more severe bottlenecks, faster reduction in local genetic diversity, and, therefore, larger values of~$D_{g}$. 

The exact solution of Eq.~(\ref{EOMH}) is given in~\citep{Korolev:Review} and is plotted in Fig.~\ref{fig:heterozygosity}A for a particular choice of the parameters. The coalescence term in Eq.~(\ref{EOMH}) acts as a sink and is responsible for the development of the dip in~$H(t,\phi)$ around~$\phi=0$. This decrease of the probability to find two different alleles at the same spatial location~[given by $H(t,0)$] tracks local fixation and the formation of sectors. 

There are two important time scales for the genetic demixing process in the model. The first time scale characterizes the interplay of migration and genetic drift and is the time over which genetic diversity is lost locally (monoallelic domains form) when radial growth is slow. As it was shown in  ~\citep{Korolev:Review}, this time scale is given by

\begin{equation}
\tau_{s}=D_{s}/D_{g}^{2}.
\label{Etaus}
\end{equation}

\noindent The other time scale characterizes the rate of the spatial expansion and is the time it takes the colony to double its initial radius:

\begin{equation}
\tau_{d}=R_{0}/v_{\parallel}.
\label{Etaud}
\end{equation}

\noindent As we shall show below, the ordering of these time scales~($\tau_{d}>\tau_{s}$ vs. $\tau_{d}<\tau_{s}$) determines whether the domain formation is sensitive to the constant increase in the front circumference.

For long times~$t\gg\tau_{d}$ and~$t\gg\tau_{s}$, the average spatial heterozygosity~$H(t,\phi)$ reaches a steady state, see Fig.~\ref{fig:heterozygosity}A and  ~\citep{Korolev:Review}. At this point, the diffusive motion of the sector boundaries is negligible compared to the deterministic increase in separation among sector boundaries due to colony expansion; therefore, the sector boundaries stop coalescing, and their relative angular positions remain largely unchanged. The number of sectors in this asymptotic state is calculated in~\citep{Korolev:Review}, and is given by

\begin{equation}
\mathcal{N}=\frac{2\pi H_{0}v_{\parallel}}{D_{g}}+H_{0}\sqrt{\frac{2\pi R_{0}v_{\parallel}}{D_{s}}}.
\label{ENR}
\end{equation}

\noindent The two terms on the right hand side of Eq.~(\ref{ENR}) reflect the dominant contributions of the two possible regimes for different orderings of~$\tau_{d}$ and~$\tau_{s}$. The second term on the right hand side of Eq.~(\ref{ENR}) gives the dominant contribution to~$\mathcal{N}(R_{0})$ when sectors form before the diffusive motion of sector boundaries becomes weaker than their deterministic separation due to colony expansion. This condition is equivalent to~$\tau_{s}\ll\tau_{d}$, i.e. the sectors form before the colony expands significantly.\footnote{A sector of angle~$\Phi$ increases in size as~$\Phi R(t)$, so the change of the sector size due to the colony growth in time~$\Delta t$ is~$\Delta t\Phi v_{\parallel}$. The change of this sector size due to the diffusive motion of the boundaries is on the order of~$\sqrt{D_{s}\Delta t}$. These two changes are equal for~$\Delta t^{*}\sim D_{s}/(v_{\parallel}^{2}\Phi^{2})$. The typical size of sectors when they first appear is about~$\sqrt{D_{s}\tau_{s}}=D_{s}/D_{g}$; therefore~$\Delta t^{*}\sim R_{0}^{2}D_{g}^{2}v_{\parallel}^{-2}D_{s}^{-1}=\tau_{d}^{2}/\tau_{s}$. Thus, for~$\tau_{s}\ll\tau_{d}$, $\Delta t^{*}\gg \tau_{d}\gg\tau_{s}$, which implies that sector boundaries appear much earlier than the time when the deterministic expansion moves them far apart. Consequently, sector boundaries have a significant chance to coalesce, eliminating some of the sectors.} This term was first calculated in  ~\citep{HallatschekNelson:ExperimentalSegregation,Hallatschek:LifeFront} by neglecting the initial stage of genetic demixing when distinct sectors appear. The first term on the right hand side of Eq.~(\ref{ENR}) is an additional contribution to~$\mathcal{N}(R_{0})$ from the early stage of genetic demixing and gives the dominant contribution in the opposite limit~$\tau_{s}\gg\tau_{d}$ when sectors do not coalesce after formation. 

Equation~(\ref{ENR}) allows us to estimate~$D_{s}$ and~$D_{g}$ (see Methods), and forms the basis of the quantitative test of spatial population genetics discussed below.

\subsection*{Model 2: Tracking domain boundaries}

Although Eq.~(\ref{EOMH}) explains spatial genetic demixing during range expansions, it does not directly describe the dynamics of the domain boundaries, one of the most prominent features of the spatio-genetic patterns shown in Fig.~\ref{fig:illustrations}. We now introduce a second modeling approach to understand genetic demixing that ignores the dynamics of individual organisms and instead focuses on a larger spatial scale of domain boundaries among monoallelic regions. This forward-in-time approach is simpler than, and complementary to, the coalescent analysis outlined in the previous section.  

After the sectors form, the dynamics of the population can be described in terms of sector boundaries. As ancestral lineages, sector boundaries diffuse and coalesce upon meeting. When domain boundaries meet, the number of domains decreases, and the average sector size increases, further increasing the average separation between the alleles. Random walks that disappear upon meeting each other (so called annihilating random walks) have been extensively studied in linear geometries~\citep{Odor:NonEquilibrium,Bramson:FiniteSize,Masser:Potts}. In the circular geometry of interest to us here, a phenomenological description of domain boundary motion consistent with Eq.~(\ref{EOMH}) is given in~\citep{HallatschekNelson:ExperimentalSegregation,Hallatschek:LifeFront} and is summarized by the following stochastic differential equation:

\begin{equation}
\label{EOMB}
d\varphi(r)=\frac{v_{\perp}}{v_{\parallel}r}dr+\sqrt{\frac{2D_{s}}{v_{\parallel}r^{2}}}dB(r),
\end{equation}

\noindent where~$\varphi(r)$ is the trajectory of a sector boundary in polar coordinates,~$B(r)$ is a Weiner process, and~$v_{\perp}$ is the velocity of the boundary perpendicular to the direction of the expansion. As in Eq.~(\ref{EOMH}), the factors of~$r$ are needed to switch from Cartesian to polar coordinates. Note that~$D_{s}$, the diffusion constant of the boundaries in Eq.~(\ref{EOMB}), is exactly the same as the diffusion constant of ancestral lineages in Eq.~(\ref{EOMH}). The equality of diffusion constants can be established by comparing the average domain sizes calculated using model 2 and model 1 in the limit~$D_{g}\rightarrow\infty$. This equality has also been derived in  ~\citep{Hallatschek:FisherWave} for the one-dimensional stepping-stone model, which describes range expansions with linear fronts.

In the model of ~\citep{HallatschekNelson:ExperimentalSegregation,Hallatschek:LifeFront}, the perpendicular velocity~$v_{\perp}$ accounts for the growth rate difference between the organisms on the opposite sides of the boundary. For neutral genetic demixing, $v_{\perp}$~equals zero, unless the colony growth is \textit{chiral}. By chiral growth we mean that sector boundaries and ancestral lineages twist exclusively clockwise or exclusively counterclockwise. This should be distinguished from sector boundary bending due to unequal fitnesses of the strains, which causes sector boundaries around the fitter strain to bend in opposite directions. 

In fact, we found that \textit{E. coli} colonies have a noticeable chirality~\footnote{Macroscopic chiral patterns are often related to microscopic symmetry breaking. After this paper was submitted for publication, we learned of work by~\cite{wang:chirality} which reveals a left-handed twisting motion of the \textit{E. coli} cell wall by tracking various markers. Given the approximate statistical bias observed by~\cite{HallatschekNelson:ExperimentalSegregation} for the first few close-packed rows of actively growing \textit{E. coli} cells to be oriented \textit{tangent} to the border of an actively growing frontier, the screw-like growth observed by Wang et al. on the micron scale might explain the macroscopic chiral patterns on a centimeter scale. Indeed, the advance of a left-handed screw at the frontier would result in sector boundaries twisting counterclockwise when viewed from the bottom of a Petri dish, as we see in Fig.~\protect{\ref{fig:boundaries}}A.}, see Fig.~\ref{fig:boundaries}A. Indications of weak chirality for the same strains were reported in~\citep{HallatschekNelson:ExperimentalSegregation}; however, chirality is more pronounced under our growth conditions. From Eq.~(\ref{EOMB}), the average twisting of the sector boundary should obey the following equation

\begin{equation}
\label{EAveragePhi}
 \mathbb{E}[\varphi(r)] = \varphi(r_{i})+ \frac{v_{\perp}}{v_{\parallel}}\ln\left(\frac{r}{r_{i}}\right),
\end{equation}

\noindent where the radius~$r_{i}$ and angle~$\varphi(r_{i})$ define the initial point from which we follow the boundary. (The subscript~$i$ in~$r_{i}$ refers to ``initial'' and is not used for indexing purposes.) The initial radius~$r_{i}$ is larger than~$R_{0}$ because the boundaries are clearly visible only after the initial stage of genetic demixing. The perpendicular velocity~$v_{\perp}$ can then be estimated from the slope of~$\varphi(r)$ vs.~$\ln(r/r_{i})$. Once the chiral parameter~$v_{\perp}$ is known, we can subtract the deterministic part from the motion of the boundary and focus on the diffusive part described by Eq.~(\ref{EOMB}) with~$v_{\perp}=0$, as we shall do in the rest of this paper. We shall further assume that~$\varphi(r_{i})=0$ since we can always rotate our reference frame by an arbitrary angle. Note that Eq.~(\protect{\ref{EOMH}}) remains valid for colonies with chiral growth like the one shown in Fig.~\protect{\ref{fig:boundaries}}A because the average spatial heterozygosity, defined as the probability of two sampled individuals to carry different alleles, depends on the separation between the two sampled individuals~(and not their absolute angular positions), which is unaffected by the deterministic twisting of the ancestral lineages. We note in passing that the average sector shape given by Eq.~(\ref{EAveragePhi}) defines an equiangular logarithmic spiral~\citep{Huntley:Divine}, with the angle determined by the dimensionless ratio~$v_{\perp}/v_{\parallel}$.

To characterize the randomness of domain boundaries motion, we compute the variance of~$\varphi(r)$ from Eq.~(\ref{EOMB}):

\begin{equation}
\label{EAveragePhi2}
 \mathrm{Var}(\varphi) = \frac{2D_{s}}{v_{\parallel}}\left(\frac{1}{r_{i}}-\frac{1}{r}\right).
\end{equation}

\noindent As~$r\rightarrow\infty$, the variance of~$\varphi$ reaches a finite limit, which is consistent with the prediction of Eq.~(\ref{EOMH}) that the coalescence of domain boundaries ceases at long times. The predicted inverse dependence of~$\mathrm{Var}(\varphi)$ on~$r$ can be used to check that domain boundaries behave as circular random walks and to estimate~$D_{s}$ from the experimental data. Indeed, $2D_{s}/v_{\parallel}$ is the slope of the line~$\mathrm{Var}(\varphi)$ vs. $1/r_{i}-1/r$. 

The annihilating random walks of sector boundaries and ancestral lineages provide an intuitive explanation of genetic demixing, but a quantitative description of genetic drift in bacterial colonies is needed to analyze evolutionary experiments done on Petri dishes as precisely as those done in liquid culture. The experiments and simulations presented here were designed to test whether the dynamics of bacterial colonies could be accurately described by Eqs.~(\ref{EOMH}) and~(\ref{EOMB}) by checking their predictions: Eqs.~(\ref{ENR}), (\ref{EAveragePhi}), and~(\ref{EAveragePhi2}).

\section*{Results}

\label{SR}

We performed experiments and on and off-lattice simulations~(see Fig.~\ref{fig:illustrations}) to test whether our models accurately describe the development of spatio-genetic correlations during range expansions. In the on-lattice simulations, range expansions are driven by migration into unoccupied territories followed by growth to saturation, similar to the range expansions of natural populations over long distances. In contrast, the off-lattice simulations were designed to mimic microbial colonies whose range expansions are driven by the mechanical forces that cells exert on each other as they grow. The utility of the simulations is threefold. First, they allow us to confirm that it is indeed possible to describe two-dimensional expansions with a one-dimensional model. Second, the simulations show that spatial genetic demixing is a generic phenomenon occurring in populations with very different biological properties and therefore growth dynamics. This in turn enables us to test whether our phenomenological model can accurately describe genetic demixing in these quite different populations in addition to our experiments. Third, the simulations allows us explore whether sector boundaries are affected by roughening of the expansion front~\citep{Hallatschek:LifeFront,Kardar:KPZ} for parameter values of interest to us here.

\begin{figure}[!ht]
\begin{center}
\includegraphics[width=88mm]{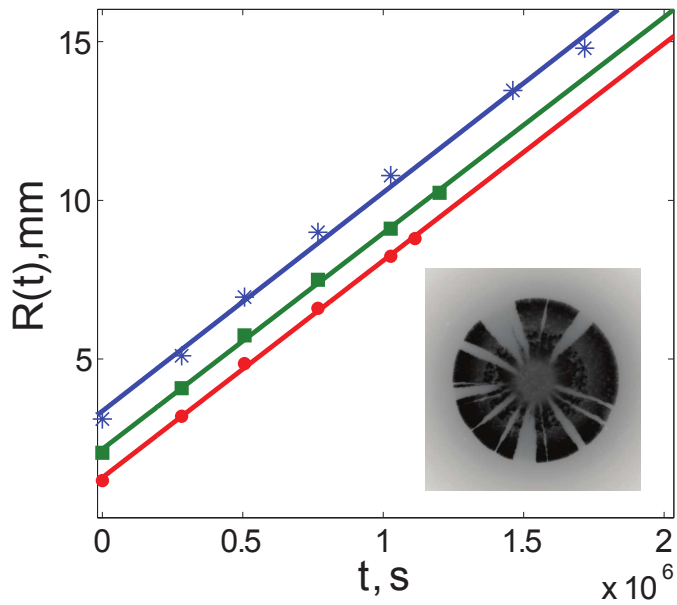}
\caption{\small Linear increase of colony radii with time. The radii of three colonies of \textit{P. aeruginosa} grown at~$21^\circ{\rm C}$ are plotted as the functions of time. The three sets of data correspond to three different inoculation volumes: $1\;\mu\rm{l}$ (red dots), $3\;\mu\rm{l}$ (green squares), and $14\;\mu\rm{l}$ (blue stars). The solid lines are the least square fits to straight lines. For all three colonies, the radius increased linearly with time and the rate of the increase was approximately independent of the initial size of the colony, as would be the case under high nutrient conditions. Note that, for the largest colony, the expansion seems to slow down slightly at the end, possibly due to nutrient depletion. Similar behavior was observed for other colonies of \textit{P. aeruginosa} and \textit{E. coli} for all growth conditions studied. The obtained values of the expansion velocities are given in Table~\protect{\ref{TNRFit}}. The inset shows a fluorescent image of \textit{P. aeruginosa} colony grown at~$21^\circ{\rm C}$; fluorescence from only one of the two alleles is shown.}
\label{fig:velocities}
\end{center}
\end{figure}

For both experiments and simulations, we found that the expansion velocity remained constant during the time of observation, see Fig.~\ref{fig:velocities}. The values of the expansion velocities are given in Table~\ref{TNRFit}. We also found that the motility and growth of the cells behind the front were significantly reduced, and the sectoring pattern did not change in the interior of the colony. Therefore, the sectoring pattern was a record of the genetic composition of the colony front throughout the expansion: A ring of radius~$r>R_{0}$ around the center of the colony represents the state of the colony front at time~$t=(r-R_{0})/v_{\parallel}$. This mapping of time onto radius allowed us to relate the time dependence of any variable in our one-dimensional model to the two-dimensional pattern in a colony.

A very direct way to verify the theory is to compare the solution of Eq.~(\ref{EOMH}) to~$H(t,\phi)$ obtained in simulations and experiments, see Fig.~\ref{fig:heterozygosity}. We found a qualitative agreement between the experiments, simulations, and the theory, but quantitative comparisons could not be made because we were unable to reliably correlate the fluorescence intensity to the fraction of the corresponding cells at a given pixel. Therefore, we focused on the two variables in the model that can be examined quantitatively: the random-walk-like motion of the sector boundaries~$\varphi(r)$ and the number of surviving sectors at long times~$\mathcal{N}(R_{0})$. For both variables, we first confirmed that the experimental data was consistent with the corresponding analytical dependence. After this, we estimated all four model parameters:~$D_{s}$, $D_{g}$, $R_{0}$, and $v_{\parallel}$, and checked that different estimates agree.

\subsection*{Lateral movement of genotypes.}

\begin{figure}[!ht]
\begin{center}
\includegraphics[width=88mm]{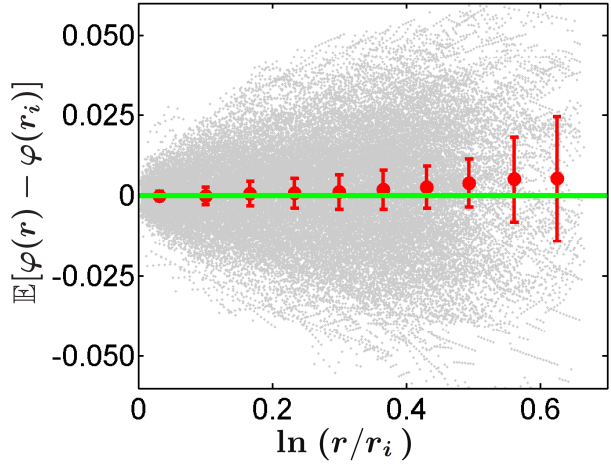}
\caption{Neutrality of \textit{P. aeruginosa} strains at~$37^{\circ}\rm{C}$. The gray dots represent individual measurements along the sector boundaries at pixel resolution (one pixel is~$50\;\mu$m) from~$24$ colonies~($524$ sector boundaries) of \textit{P.~aeruginosa}. Positive~$\varphi$ represent boundary bending from cyan towards yellow sectors. Each red dot is the mean position of the gray dots in one of the~$10$ equidistant bins along the x-axis. The error bars represents~$95$\% confidence interval of the mean values. The green line is the expected dependence if the strains are neutral,~$\mathbb{E}[\varphi(r)]=0$. Since the expected line passes through all of the error bars, we cannot reject the hypothesis that the strains are neutral. In fact, the largest fitness difference consistent with the data is about~$10^{-4}$. This comes from using Eq.~(\protect{\ref{EAveragePhi}}) to find~$v_{\perp}/v_{\parallel}$ and relating~$v_{\perp}/v_{\parallel}$ to the fitness difference as in~\protect{\citep{Hallatschek:LifeFront,korolev:thesis}}. It is then reasonable to conclude that these strains can be considered equally fit on the time scales of our experiments. The corresponding data for \textit{P.~aeruginosa} strains grown at~$21^{\circ}\rm{C}$ is also consistent with the assumption that the strains are equally fit. \textit{E. coli} strains are also equally fit, see Fig.~\protect{\ref{fig:boundaries}} and~\protect{\citep{HallatschekNelson:ExperimentalSegregation}}.}
\label{fig:neutrality}
\end{center}
\end{figure}

\begin{figure}[!ht]
\begin{center}
\includegraphics[width=180mm]{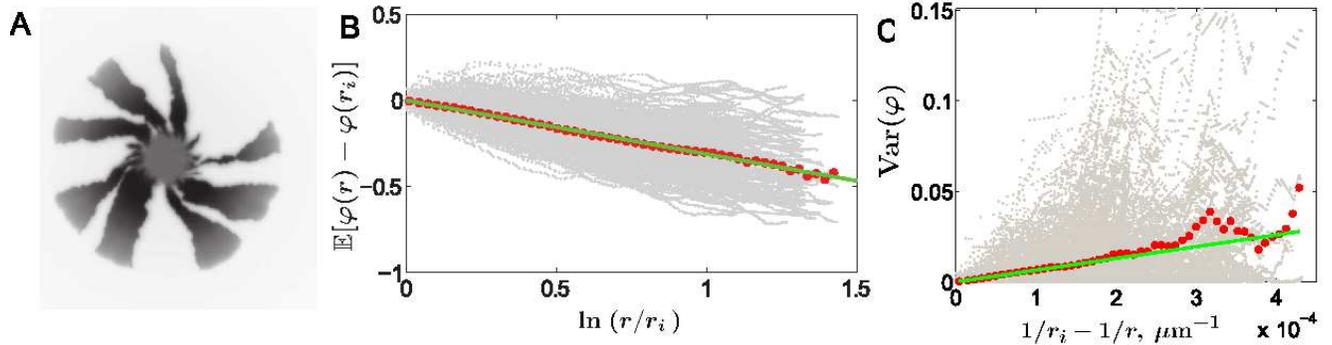}
\caption{Chirality and boundary wandering in \textit{E. coli} colonies. (A) Genetic demixing in a chiral \textit{E. coli} colony. The image, taken from the bottom of the Petri dish, shows the fluorescent signal from only one of the two segregating alleles. Bacterial colonies of \textit{P.~aeruginosa} at~$37^\circ{\rm C}$ and~$21^\circ{\rm C}$ did not exhibit chiral growth. (B) Twisting of sector boundaries in \textit{E. coli} colonies. The gray dots represent individual measurements along the sector boundaries at pixel resolution (one pixel is~$50\;\mu$m) from~$30$ colonies~($390$ sector boundaries). Each red dot is the mean position of the gray dots in one of the~$50$ equidistant bins along the x-axis. The green line is the least square fit to the red dots. According to Eq.~(\ref{EAveragePhi}), the slope of the green line equals~$v_{\perp}/v_{\parallel}$, which yields~$v_{\perp}/v_{\parallel}=0.32$. As shown in (A) and (B), all sector boundaries twist on average in the same direction for chiral growth. In contrast, sector boundaries between non-neutral strains bend in both clockwise and counterclockwise directions because the boundaries around the fitter strains bend outwards~\protect{\citep{Hallatschek:LifeFront}}. To exclude possible small fitness differences, we also separately analyzed boundaries that would bend in opposite directions if the strains were non-neutral. Non-neutrality would result in different values of~$v_{\perp}/v_{\parallel}$ for these two sets of boundaries. Upon applying equivalent tests to those used in Fig.~\protect{\ref{fig:neutrality}} we confirmed that~$v_{\perp}/v_{\parallel}$ are the same, which supports the previous finding that these strains are neutral~\protect{\citep{HallatschekNelson:ExperimentalSegregation}}.  (C) Random walks of sector boundaries. The same data as in (B) is used to plot the variance of~$\varphi(r)$; gray dots in the individual data points, and the red dots are the averages. The green line is the least square fit to the first~$25$ red dots; the last~$25$ dots are not used because of large fluctuations due to a smaller sample size at large~$r$. According to Eq.~(\protect{\ref{EAveragePhi2}}), the slope of the green line equals~$2D_{s}/v_{\parallel}$. The data sets for \textit{P.~aeruginosa} at~$37^\circ{\rm C}$ and~$21^\circ{\rm C}$ show similar behavior of the variance of~$\varphi(r)$, but the data set for \textit{E.~coli} fluctuates less partly because of the larger sample size and larger spatial diffusion constant compared to the other two experiments.}
\label{fig:boundaries}
\end{center}
\end{figure}

By tracing a large number of sector boundaries~$\varphi(r)$, we verified that their trajectories had the statistical properties of random walks in the circular geometry. For \textit{P. aeruginosa} colonies, the average angular positions of sector boundaries remained constant indicating equal fitness of the strains; see Fig.~\ref{fig:neutrality}. \textit{E.~coli} colonies experienced chiral growth, and the boundaries twisted counterclockwise on average; see Fig.~\ref{fig:boundaries}A. This twisting [also observed in~\citep{HallatschekNelson:ExperimentalSegregation}] was consistent with constant lateral velocity~$v_{\perp}$ and Eq.~(\ref{EAveragePhi}) as shown in Fig.~\ref{fig:boundaries}B. We also analyzed the deviations of sector boundaries from their average position. As shown in Fig.~\ref{fig:boundaries}C, the variance of~$\varphi(r)$ was consistent with the random walk behavior in the circular geometry summarized by Eq.~(\ref{EAveragePhi2}).

\subsection*{Genetic diversity at the population front.}

\begin{figure}[!ht]
\begin{center}
\includegraphics[width=15.2 cm]{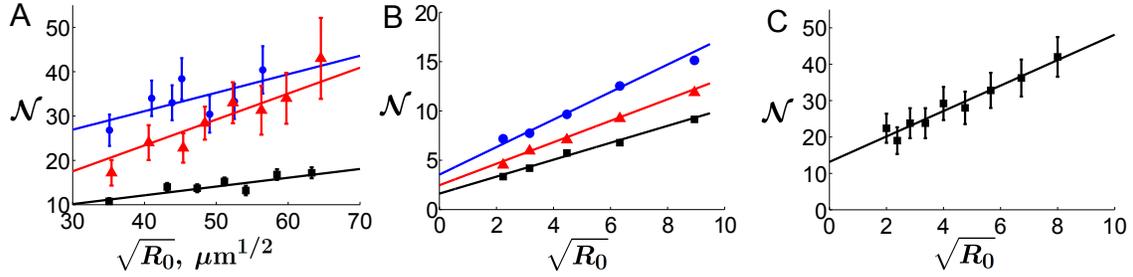}
\caption{\small Dependence of the diversity at the expansion front on the initial size of the population in experiments~(A), on-lattice simulations~(B), and off-lattice simulations~(C). The average number of sectors at the end of a range expansion~$\mathcal{N}$ is plotted versus the square root of the initial radius of the colony~$R_{0}$. These coordinates are chosen so that the theoretically predicted dependence~[see Eq.~(\protect{\ref{ENR}})] is a straight line~(solid lines). The error bars represent 95\% confidence interval. Parameters~$D_{s}/v_{\parallel}$ and~$D_{g}/v_{\parallel}$ can be estimated from the slope of the fit and its intercept with the y-axis respectively. (A) \textit{E. coli}~(black squares), \textit{P. aeruginosa} at~$37^\circ{\rm C}$ (red triangles), and \textit{P. aeruginosa} at~$21^\circ{\rm C}$ (blue dots). (B) Three different island carrying capacities: $N=3$ (black squares), $N=30$ (red triangles), and $N=300$ (blue dots). Each data point is the average of 40 simulations. Notice that the population with higher~$N$ have more sectors in agreement with Eq.~(3) because populations with higher densities have lower~$D_{g}$. In this plot, the error bars are represented by the size of the markers used. (C) Due to the discrete nature of these simulations, we use the square root of the number of cells at inoculation as a proxy for the initial radius.}
\label{fig:sectors}
\end{center}
\end{figure}

Our model of genetic demixing during bacterial range expansions relies on three assumptions: equal fitness of the strains, expansion with a constant velocity, and diffusive migrations at the colony front. Since these assumptions were confirmed by the experimental data, we turned to the number of surviving sectors given by Eq.~(\ref{ENR}), which relies not only on the aforementioned assumptions but also on the solution of the model. We found that our one-dimensional model provided a good description of two-dimensional simulations (Fig.~\ref{fig:sectors}B and C). The experimentally observed~$\mathcal{N}(R_{0})$ was also consistent with the predicted square-root dependence for both \textit{E. coli} and \textit{P. aeruginosa}, as one can see in Fig.~\ref{fig:sectors}A.  

\subsection*{Quantitative test of the models.}

The agreement between the experiments and the theory not only suggests that bacterial expansions can be described by the one-dimensional stepping-stone model, but also allows us to put this agreement to a quantitative test.

From the fit of Eq.~(\ref{ENR}) to the dependence of sector number on the initial population radius, we estimate the two key parameters from the model~$D_{s}/v_{\parallel}$ and~$D_{g}/v_{\parallel}$. The estimate of~$D_{s}/v_{\parallel}$--the degree of lateral movement by the genotypes as the population expands--forms a prediction that allows us to test the ability of Eq.~(\ref{EOMH}) to capture the process of population expansion and genetic drift in the bacteria. This test is performed by taking the predicted value of~$D_{s}/v_{\parallel}$ and comparing it to a direct estimate that is made using our measurements of the sector boundary trajectories. Specifically, the diffusion constant of random walks performed by sector boundaries was calculated by fitting the observed variance of~$\varphi(r)$ to the expected dependence for a random walk in the circular geometry given by Eq.~(\ref{EAveragePhi2}), see Fig.~\ref{fig:boundaries}C. The parameters estimated by the two methods are summarized in Table~\ref{TNRFit}. The quantitative agreement between these quite different estimates of~$D_{s}/v_{\parallel}$ suggests that our spatial population genetics model can indeed describe the simple bacterial range expansions discussed here.

Although it is beyond the scope of this paper to quantitatively relate the microscopic properties of cells and colonies to the macroscopic parameters of our phenomenological model, we can nevertheless understand how to infer some information about the processes within the colony from the measured values of~$D_{s}$ and~$D_{g}$. To make this connection, we need to know the cell size~$a_{c}$ and the generation time~$\tau$. These parameters do not have well-defined values because cells are rod shaped and the generation time presumably depends on the location of a cell relative to the front, where the nutrients are readily available. Therefore, we did not measure~$a$ and~$\tau$, but we believe that it is reasonable to assume that $a_{c}$ is on the order of~$1-5\;\mu$m, and~$\tau$ is on the order of~$10^{3}-10^{4}\;$s~\citep{yang:generation_lung,bartlett:size}. Given the wide range of uncertainty and the differences in~$a_{c}$ and~$\tau$ between different experimental conditions, we can only make a very rough estimates, and the numbers given below should be considered accurate within an order of magnitude. The velocity of expansion is given by~$N_{v}a_{c}/\tau$, where~$N_{v}$ is the number of actively growing cells at the front~\citep{HallatschekNelson:ExperimentalSegregation}. From Table~\ref{TNRFit},~$N_{v}$ is on the order of ten cells, which is similar to the earlier results~\citep{HallatschekNelson:ExperimentalSegregation}. The square root of the product of the diffusion constant and the generation time is proportional to the root mean square lateral displacement of a cell during one generation. From the data, we estimate that this lateral displacement is on the order of ten cell sizes, which comparable to the advancement of the front during the same period of time. As we show in Appendix A [Eq.~(\ref{eq:Dg})], the strength of genetic drift~$D_{g}$ can be expressed in terms of effective population density~$\rho_{e}$; in fact,~$D_{g}=1/(\rho_{e}\tau)$. We find that the effective number of cells in a spatial region of the size of a cell,~$a_{c}$, is smaller or much smaller than one, as has been predicted by~\cite{Hallatschek:1dWave}.

\section*{Conclusions}
\label{SC}

Expanding bacterial colonies are subject to strong genetic drift, resulting in local extinctions of the genotypes present in the population. These local extinctions manifest themselves in the formation of mono-allelic sectors, provided the organisms migrate very slowly behind the expansion front. Despite the potential for complexity in the physical and biotic interactions within bacterial groups, our experiments strongly suggest that these boundaries behave as annihilating random walks, and we show how to estimate their diffusion coefficients from the fluorescent images of spatial genetic demixing (see Methods).

The spatial separation of two alleles in an expanding bacterial colony can also be described by a generalization of the stepping-stone model, which accounts not only for the behavior of the sector boundaries but also for their formation from the initially well-mixed population. Although the expansion takes place on a two-dimensional surface, our experiments and simulations strongly suggest that its effects on the genetic composition of the population front can be described by a simple one-dimensional model with only four parameters ($D_{s}$, $D_{g}$, $R_{0}$, and $v_{\parallel}$), all of which can be measured experimentally as described in Methods.

This generalization of the stepping-stone model not only explains spatial genetic demixing but also characterizes neutral evolutionary dynamics of expanding populations quantitatively, just as the Wright-Fisher and Moran models capture essential aspects of the dynamics of stationary well-mixed populations. This quantitative description and the methods of estimating the model parameters could be important for understanding evolution within microbial communities. For example, the survival probability of a beneficial mutation depends on~$D_{g}$, characterizing the strength of genetic drift in spatially extended populations~\citep{Maruyama:Fixation,Doering:FisherWaveWeakSelection,Korolev:Review}, and the evolution of cooperation depends on the relatedness~(or assortment) of the organisms, which is described by~$H(t,\phi)$~\citep{Nadell:EmergenceSpatialStructure,Nadell:Sociobiology}. In addition, a better understanding of bacterial range expansions could also facilitate the development of new methods of genetic inference from spatially resolved data. 

More generally, we believe that a key contribution of this study is to illustrate the potential of spatial population genetic models to make quantitative predictions. Clearly, our bacterial system is not representative of all species and is a system that lends itself to making measurements that will not always be possible in other species. Nevertheless, our study demonstrates the potential for these models to generate accurate quantitative predictions as well as confirming the utility of population genetics as a method for evolutionary analysis.

\section*{Acknowledgments}
D.R.N. and K.S.K. are grateful to Oskar Hallatschek for many helpful conversations and for a critical reading of the manuscript. Some of the computations in this paper were run on the Odyssey cluster supported by the FAS Sciences Division Research Computing Group.

This work was supported by a National Institute of General Medical Sciences Center of Excellence grant (5P50 GM 068763-01) to K.R.F. and to D.R.N., and European Research Council Grant 242670 to K.R.F.. D.R.N. also acknowledges the support of the National Science Foundation through Grant DMR-0654191 and through the Harvard Materials Research Science and Engineering Center through Grant DMR-0820484.

\appendix

\section*{Appendix A: Genetic drift in expanding populations}

In this Appendix, we briefly outline the theory leading to Eq.~(\ref{EOMH}) and discuss the relationship between~$D_{g}$ and effective population size~$N_{e}$. Because the number of offspring fluctuates, allele frequencies change stochastically. In the classic Wright-Fisher model of a haploid population with two neutral alleles at a locus, the dynamics of the frequency of one of the alleles,~$f(t)$, is governed by the following stochastic differential equation

\begin{equation}
\label{eq:f_well_mixed}
df(t)=\sqrt{\frac{1}{N\tau}f(t)[1-f(t)]}dB(t),
\end{equation}

\noindent where~$B(t)$ is a Weiner process, and~$N$ is the population size. The generation time~$\tau$ is shown explicitly to facilitate the transition to the spatial model later. For other population models, e.g. the Moran model, Eq.~(\ref{eq:f_well_mixed}) still describes the dynamics of~$f(t)$, but with a different coefficient inside the square root~\citep{Kingman:Coalescent,Wakeley:Coalescent}. It is then convenient to introduce an effective population size~$N_{e}$ instead of~$N$ so that Eq.~(\ref{eq:f_well_mixed}) is satisfied. Under certain assumptions, the effective population size equals the ratio of census population size to the variance in the number of offspring~\citep{Kingman:Coalescent}. This choice of~$N_{e}$ is sometimes called coalescent effective population size~\citep{Wakeley:Coalescent} or eigenvalue effective population size~\citep{ewens:ne}.

From stochastic differential equation~(\ref{eq:f_well_mixed}), we can compute moments (e.g., average or variance) using It\^{o}'s lemma~\citep{Risken:FPE,Gardiner:Handbook}. For example, the average heterozygosity~$H(t)=\mathbb{E}[2f(t)[1-f(t)]$ obeys the following equation

\begin{equation}
\label{eq:H_derive}
\frac{d}{dt}H(t)=-\frac{1}{N_{e}\tau}H(t),
\end{equation}

\noindent describing the loss of genetic diversity due to genetic drift.

The one-dimensional stepping-stone model is a generalization of Eq.~(\ref{eq:f_well_mixed}) to an array of demes labeled by an index~$k$~\citep{KimuraWeiss:SSM,Barton:NeutralEvolution}:

\begin{equation}
\label{eq:f_ssm}
df_{k}(t)=\frac{m}{2\tau}[f_{k-1}(t)-2f_{k}(t)+f_{k+1}(t)]+\sqrt{\frac{1}{N_{e}\tau}f_{k}(t)[1-f_{k}(t)]}dB_{k}(t),
\end{equation}

\noindent where~$m$ is the fraction of organisms migrating out of each deme into one of the two nearest neighbors. We can obtain a continuous in space description of this lattice model by introducing a spatial variable~$x=ka$, where~$a$ is the distance between the demes. The corresponding stochastic differential equation reads~\citep{Korolev:Review}

\begin{equation}
\label{eq:f_ssm_continuous}
df(t,x)=D_{s}\frac{\partial^{2}}{\partial x^{2}}f(t,x)dt+\sqrt{D_{g}f(t,x)[1-f(t,x)]}dB(t,x),
\end{equation}

\noindent where

\begin{equation}
\label{eq:Ds}
D_{s}=\frac{ma^{2}}{2\tau},
\end{equation}

\noindent and

\begin{equation}
\label{eq:Dg}
D_{g}=\frac{a}{N_{e}\tau}.
\end{equation}

\noindent Note that the factor of~$a$ in the definition of~$D_{g}$ arises from the normalization of the spatial Wiener process~$B(t,x)$. The ratio~$\rho_{e}=N_{e}/a$ is the effective population density, which appears in the continuous formulations of the stepping-stone model~\citep{Nagylaki:DecayGeneticVariability}. When Eq.~(\ref{eq:f_ssm_continuous}) is applied not to strictly one-dimensional populations, but to linear expansion fronts,~$D_{g}$ depends on many demographic parameters in an intricate way~\citep{Hallatschek:1dWave}.

For circular expansions of interest to us here, it is convenient to introduce polar coordinates~$R(t)=R_{0}+v_{\parallel}t$ and~$\phi$ around the center of the expansion. Equation~(\ref{eq:f_ssm_continuous}) then takes the following form

\begin{equation}
\label{eq:f_ssm_radial}
df(t,\phi)=\frac{D_{s}}{R^{2}(t)}\frac{\partial^{2}}{\partial \phi^{2}}f(t,\phi)dt+\sqrt{\frac{D_{g}}{R(t)}f(t,\phi)[1-f(t,\phi)]}dB(t,\phi).
\end{equation}

\noindent To obtain Eq.~(\ref{EOMH}), we use It\^{o}'s lemma to differentiate~$H(t,x)$ defined by Eq.~(\ref{eq:heterozygosity}) in terms of~$f(t,\phi)$.

\newpage

\begin{table}[!ht]
\caption{Measured values of the model parameters}
\begin{tabular}{lllll}
\hline
Experiment  &  $v_{\parallel}$~($\mu{\rm m/s}$) & $D_{g}/v_{\parallel}$ & $D^{e}_{s}/v_{\parallel}$~($\mu$m) & $D^{m}_{s}/v_{\parallel}$~($\mu$m) \\ \hline

\textit{E.~coli}  & $(8.3\pm0.4)\cdot10^{-3}$ & $0.75$, $(0.32,\infty)$ & $40$, $(10,90)$  & $32\pm2$ \\

\textit{P.~aeruginosa}  at~$37^\circ{\rm C}$  & $(17.3\pm0.9)\cdot10^{-3}$ & $\infty$, $(3,\infty)$ & $4.6$, $(4.0,5.2)$ & $3.7\pm1.1$ \\

\textit{P.~aeruginosa} at~$21^\circ{\rm C}$ & $(6.9\pm0.3)\cdot10^{-3}$ & $0.17$, $(0.10,0.77)$ & $5.5$, $(1.2,13.0)$ & $5.2\pm0.5$ \\ \hline
\end{tabular}
\begin{flushleft}
{Expansion velocities, migration rates, and the strengths of genetic drift. $D^{e}_{s}$ and $D^{m}_{s}$ are two different estimates of the diffusion constant; their agreement indicates quantitative validity of our model. The values of~$v_{\parallel}$ were obtained by fitting~$R(t)$ to a linear function, see Fig.~\protect{\ref{fig:velocities}}. $D_{g}$ and~$D^{e}_{s}$ estimated from the fit of the data shown in Fig.~\protect{\ref{fig:sectors}} to Eq.~(\protect{\ref{ENR}}). For~$D_{g}/v_{\parallel}$ and~$D^{e}_{s}/v_{\parallel}$, the 95\% confidence intervals are given in parentheses. The infinite values of~$D_{g}/v_{\parallel}$ are due to the possibility of the linear fit~(see Fig.~\protect{\ref{fig:sectors}}A) passing through the origin. $D^{m}_{s}$ are measured from the random walks of sector boundaries~[see Eq.~(\protect{\ref{EAveragePhi2}})], and the error estimates represent the standard deviation between different repetitions of the experiments. See Methods for a detailed description of the estimating procedures.}
\end{flushleft}
\label{TNRFit}
\end{table}

\clearpage

\clearpage

\clearpage

\clearpage

\clearpage

\clearpage

\end{document}